\def\rfr#1{eq. (\ref{#1})}
\def\rfrs#1#2{(\ref{#1})-(\ref{#2})}

\def\Rfr#1{Eq. (\ref{#1})}

\def\dert#1#2{\frac{{{d}}{#1}}{{{d}}{#2}}}              

\def\bar{\begin{eqnarray}}
\def\ear{\end{eqnarray}}

\def\eqi{\begin{equation}}
\def\eqf{\end{equation}}
\def\eqia{\begin{eqnarray}}
\def\eqfa{\end{eqnarray}}
\def\rp#1#2{{#1\over#2}}

\def\lb#1{\label{#1}}

\def\oc2{$\mathcal{O}(c^{-2})$}

\documentclass[showpacs,twocolumn,showkeys]{revtex4}

\usepackage{amsmath,amsthm,amscd,amssymb}
\usepackage{latexsym}
\usepackage{graphicx,epsfig}
\usepackage{amsfonts}

\begin{document}
\title{Solar System planetary orbital motions and $f(R)$ Theories of Gravity}

\author{Matteo Luca Ruggiero}
\email{matteo.ruggiero@polito.it}
 \affiliation{Dipartimento di Fisica, Politecnico di Torino, Corso Duca degli Abruzzi 23, Torino, Italy\\
 INFN, Sezione di Torino, Via Pietro Giuria 1, Torino, Italy}

\author{Lorenzo Iorio}
\email{lorenzo.iorio@libero.it} \affiliation{Viale Unit\`{a} d'Italia 68, 70125 Bari, Italy}


\begin{abstract}
In this paper we  study the effects of $f(R)$  Theories of Gravity on Solar System gravitational tests. In
particular, starting from an exact solution of the field equation in vacuum, in the Palatini formalism, we work
out the effects that the modifications  to the Newtonian potential  would induce on the Keplerian orbital
elements of the Solar System planets, and compare them with the latest results in planetary orbit determination
from the EPM2004 ephemerides. It turns out that the longitudes of perihelia  and the mean longitudes are
affected by secular precessions. We obtain the resulting best estimate of the parameter $k$ which, being simply
related to the scalar curvature,  measures the non linearity of the gravitational theory. We use our results to
constrain  the cosmological constant and show how $f(R)$ functions can be constrained, in principle. What we
obtain suggests that, in agreement with other recent papers, the Solar System experiments are not effective to
set such constraints, if compared to the cosmologically relevant values.
\end{abstract}



\maketitle

\section{Introduction}\label{sec:intro}

In order to make General Relativity (GR) agree  with the recent observations in Cosmology, which evidence the
acceleration of the Universe thanks to experimental data coming from different tests \cite{c1,c2,c3,c4,c5}, the
existence of the \textit{dark energy} is needed. On the other hand, theories alternative to GR, such as Higher
Order Theories of Gravity (HOTG) can explain the acceleration of the Universe without requiring the existence of
dark energy \cite{c6,c7,c8,c9}. The simplest model of HOTG are the so-called $f(R)$ theories, where the
Lagrangian depends on an arbitrary analytic function $f$ of the scalar curvature $R$. As it is well known,
$f(R)$ theories can be studied both in the metric formalism \cite{metfr1,metfr2,metfr3,metfr4} and in the
Palatini formalism \cite{palfR1,palfR2,palfR3,palfR4,palfR5,palfR6,palfR7,palfR9}.  We remind that, in the
vacuum case, $f(R)$ theories, in the Palatini formalism, are equivalent to GR with a (non dynamical)
cosmological constant.

Since GR is in sharp agreement with the experimental results obtained in the Solar System \cite{solar}, every
theory that aims at agreeing with experimental results at cosmological scale, should reproduce GR at the Solar
System scale.

So, the problem arises of testing the reliability of these theories with Solar System experiments. The dynamical
equivalence between $f(R)$  and scalar-tensor theories of gravity (see for instance \cite{sotoriu06} and
references therein) has been used to set bounds on the analytic form of the functions $f(R)$, thanks to
post-Newtonian parameters of scalar-tensor gravity \cite{troisi05}. However, this approach was recently
criticized \cite{faraoni06}, since the dynamical equivalence has been used beyond its range of validity.
Actually, the debate is still open, and the problem can be faced both in the metric formalism and in the
Palatini formalism (see \cite{allemandi05,sotoriu05} and references therein).

Also the issue of the dynamics of $f(R)$ theories in both formalisms in the presence of matter (i.e. inside the
sources of the gravitational field) is under debate \cite{chiba03,flanagan03}. In particular, a recent paper
\cite{erick06a} states that $f(R)$ theories where a $1/R$ term is added to the Einstein-Hilbert action do not
match Solar System experiments: these conclusions have been subsequently generalized to other forms of $f(R)$
\cite{jin06}. Indeed, these analyses are carried out in the metric formalism and lead to the same results
obtained by Chiba \cite{chiba03}. The difficulties arising in the metric formalism with the Solar System
experiments, which ultimately depend on the matching between the solutions  inside and outside the matter
distribution (i.e. the matching between the star interior and the vacuum exterior), are not present in the
Palatini formalism, as it has been recently showed \cite{kainulainen06}:  in the Palatini formalism of
the $f(R)$ theories, the space-time inside a star does not affect the space-time outside it, (i.e.  the vacuum solution), contrary
to what happens in the metric formalism, even though in this approach the mass of the star and its density have
a non standard relation.

In a previous paper \cite{allemandi05}, one of us studied the consistence of $f(R)$ theories with observational
data in the Palatini formalism. We aimed at understanding the corrections to GR arising from specific
modifications of the Hilbert-Einstein Lagrangian. We found  an exact solution (corresponding to constant scalar
curvature $R$) to the field equations in vacuum and we showed that the modifications to standard GR are directly
related to the solutions of what we called \textit{the structural equation}. The latter is an algebraic
scalar-valued equation, that controls the solutions of the field equations of the theory.  These modifications
can be suitably interpreted as post-Newtonian parameters related to the non linearity of the theory.

The exact solution we found in \cite {allemandi05} corresponds to the Schwarzschild-de Sitter metric, which has
been studied in the past \cite{islam83,lake02,mashhoon03} and, also,  more recently  in connection with the
problem of the cosmological constant \cite{kagramanova06,sereno06a,sereno06b}. The relevance of the
Schwarzschild-de Sitter metric on $f(R)$ theories of gravity has also been discussed in \cite{multamaki06}.
Here, we explicitly work out the effects of $f(R)$ theories on the Keplerian orbital elements of the Solar
System planets, and compare them with the latest results in planetary orbit determination from the EPM2004
ephemerides, in order to obtain the best estimate for the parameter $k$, which is a measure of the non linearity
of the theory. We compare this estimate to the values of the cosmological constant $\Lambda$ and suggest that it
can be used to set bounds on the parameters of the $f(R)$ functions relevant in cosmology.

The paper is organized as follows: after briefly introducing the theoretical framework of $f(R)$ gravity and the
vacuum exact solution in Sec. \ref{sec:frgravity}, we investigate the perturbations of the orbital elements in
Sec. \ref{sec:planorb}, which are compared to recent data in Sec. \ref{sec:conf}; then,  we show how our
approach may lead to constraints  in cosmology in  Sec. \ref{sec:disc}. Conclusions are outlined in Sec. \ref{sec:conc}.

\section{Vacuum exact solution of $f(R)$ gravity field equations}\label{sec:frgravity}

The equations of motion of the $f(R)$ gravity in the Palatini formalism can be obtained, by independent
variations with respect to the metric and the connection,  from the action \footnote{Let the signature of the
$4$-dimensional Lorentzian manifold $M$ be $(-,+,+,+)$; furthermore,  if not otherwise stated, we use units such
that $G=c=1$.}
\begin{equation}
A=A_{\mathrm{grav}}+A_{\mathrm{mat}}=\int [ \sqrt{g} f (R)+2\chi L_{\mathrm{mat}} (\psi, \nabla \psi) ]  \;
d^{4}x, \label{eq:actionf(R)}
\end{equation}
where $R\equiv R( g,\Gamma) =g^{\alpha\beta}R_{\alpha \beta}(\Gamma )$, $R_{\mu \nu }(\Gamma )$ is the
Ricci-like tensor of any torsionless connection $\Gamma$   independent from the metric  $g$, which is assumed
here to be the physical metric. The gravitational part of the Lagrangian is represented by any real analytic
function $f (R)$ of  the scalar curvature $R$. The total Lagrangian contains also a first order matter part
$L_{\mathrm{mat}}$ functionally depending on yet unspecified  matter fields $\Psi$ together with their first
derivatives,
equipped with a gravitational coupling constant $\chi=\frac{8\pi G}{c^4}$ (see e.g. \cite{buchdahl1,buchdahl2}).\\
According to the Palatini formalism
\cite{Barraco,palfR1,palfR2,palfR3,palfR4,palfR5,palfR6,palfR7,palfR9,FFVa,FFVb}, from (\ref{eq:actionf(R)}) we
obtain the following equations of motion:
\begin{eqnarray}
f^{\prime }(R) R_{(\mu\nu)}(\Gamma)-\frac{1}{2} f(R)  g_{\mu \nu
}&=&\chi T_{\mu \nu }^{mat},  \label{ffv1}\\
\nabla _{\alpha }^{\Gamma }[ \sqrt{g} f^\prime (R) g^{\mu \nu })&=&0, \label{ffv2}
\end{eqnarray}
where $T^{\mu\nu}_{mat}=-\frac{2}{\sqrt g}\frac{\delta L_{\mathrm{mat}}}{\delta g_{\mu\nu}}$ denotes the matter
source stress-energy tensor and $\nabla^{\Gamma}$ means covariant derivative with respect to the connection
$\Gamma$. Actually, it is possible to show \cite{FFVa,FFVb} that the manifold $M$, which is the model of the
space-time, can be a posteriori endowed with a bi-metric structure $(M,g,h)$  equivalent to the original
metric-affine structure $(M,g,\Gamma)$, where $\Gamma$ is assumed to be the Levi-Civita connection of $h$. The
two metrics are conformally related by
\begin{equation}\label{h_met2}
h_{\mu \nu }=f^\prime (R)  \;  g_{\mu \nu }.
\end{equation}
The equation of motion (\ref{ffv1}) can be supplemented by the scalar-valued equation obtained by taking the
$g$-trace of (\ref{ffv1}), where we set $\tau=\mathrm{tr} T=g^{\mu \nu }T^{mat}_{\mu \nu }$:
\begin{equation}
f^{\prime} (R) R-2 f(R)= \chi \tau.  \label{ss}
\end{equation}
The algebraic equation (\ref{ss}) is called the \textit{structural equation} and it controls the solutions of equation (\ref{ffv1}).\\

The field equations (\ref{ffv1}-\ref{ffv2}) and the structural
equation (\ref{ss}) in vacuum become
\begin{eqnarray}
[f'(R)] R_{(\mu\nu)}(\Gamma)-\frac{1}{2}[f(R)] g_{\mu \nu
}&=&0 , \label{ffv111}\\
\nabla _{\alpha }^{\Gamma }(\sqrt{g} \; [f'(R)] \; g^{\mu \nu
})&=&0, \label{ffv211} \\
f^{\prime }(R) R-2f(R)&=& 0, \label{eq:fvac2}
\end{eqnarray}
As shown in \cite{allemandi05} (see, in particular, Section 3),
the system of equations (\ref{ffv111}-\ref{eq:fvac2}) has the
spherical symmetrical solution
\begin{align}
ds^2=&-\left(1-\frac{2m}{r}+\frac{k
r^2}{3}\right)dt^2+\frac{dr^2}{\left(1-\frac{2m}{r}+\frac{k
r^2}{3}\right)} \notag \\ & +r^2d\vartheta^2+r^2\sin^2 \vartheta
d\varphi^2, \label{eq:metrica1}
\end{align}
where $m$ is the mass of the  source of the gravitational field
and $k=c_i/4$, where $R=c_i$ is any of the solutions of the
structural equation (\ref{eq:fvac2}). In doing so, we have
obtained a solution with constant scalar curvature $R$.

In particular, if $f(R)=R$, i.e. our theory is GR, $R=0$ is the
solution of the structural equation, and (\ref{eq:metrica1})
reduces to the classical Schwarzschild solution.  Again, for a
thorough discussion we refer to \cite{allemandi05}, and references
therein.

The metric (\ref{eq:metrica1}) corresponds to the Schwarzschild-de
Sitter space-time (see \cite{he,ruffi}), which is exactly a
solution of standard GR with fixed value cosmological constant.

From (\ref{eq:metrica1}) it is evident that the modifications to the standard Schwarzschild solution of vacuum
GR are proportional to the values of the Ricci scalar, owing to the proportionality between $k$ and $c_i$.
Indeed, this contribution to the gravitational potential should be small enough not to contradict the known
tests of gravity. In the cases of small values of $R$ (which surely occur at Solar System scale) the Einsteinian
limit (i.e. the Schwarzschild solution) and
 the Newtonian limit are recovered, as it is evident from
(\ref{eq:metrica1}), and it has been proved in \cite{allemandi05}. This implies, also, that in order to study
the corrections deriving from $f(R)$ theories on the orbital elements in the Solar System, the terms
proportional to $k$ can be treated as a perturbation.

For further convenience, we remark that, thanks to the following change of the radial variable
\begin{equation}
\overline{r}  = r \left(1-\frac{m}{r}-\frac{k r^2}{12} \right),  \label{eq:iso6}
\end{equation}
the metric (\ref{eq:metrica1}) can be written in isotropic form
\begin{align}
ds^2=&-\left(1-\frac{2m}{\overline r}+\frac{k \overline r^2}{3}\right)dt^2 \notag \\ &+
\left(1+\frac{2m}{\overline r} +\frac{k \overline r^2}{6} \right)\left(d\overline r^2+\overline r^2
d\vartheta^2+\overline r^2 \sin^2 \vartheta d\varphi^2\right), \label{eq:iso8}
\end{align}
up to first order in $m$ and $k$.

\section{The impact of $f(R)$ gravity  on the planetary orbits}
\label{sec:planorb}

According to what we have seen before, the modifications to the solutions of the field equations due to $f(R)$
theories are given by a term proportional to the Ricci scalar. As a consequence, from (\ref{eq:iso8}) we can
consider a perturbation of the gravitational potential in the form \eqi \Delta U=\kappa r^2\lb{poti},  \eqf
where $\kappa=k/3$. In doing so, we neglect the effect of spatial curvature.

From the potential (\ref{poti}) we obtain an entirely radial acceleration \eqi {\boldsymbol{A}}=-2\kappa
\boldsymbol{r}.\lb{ADM}\eqf Its effect on planetary motions, which is, of course, much smaller than usual
Newtonian gravity, can straightforwardly be calculated within the usual perturbative schemes (see, for instance,
\cite{roy}), i.e. using the Gauss equations, which enable us to study the perturbations of the Keplerian
elements, induced by generic perturbing accelerations. Namely, the Gauss equations for the variations of the
semimajor axis $a$, the eccentricity $e$, the inclination $i$, the longitude of the ascending node $\Omega$, the
argument of pericentre $\omega$ and the mean anomaly $\mathcal{M}$ of a test particle in the gravitational field
of a body $m$ are, in general, given by

\begin{widetext}
\bar
\dert{a}{t} & = & \rp{2}{n\sqrt{1-e^2}}\left[eA_r\sin \varphi+A_t\left(\rp{p}{r}\right)\right],\lb{smax}\\
\dert{e}{t} & = & \rp{\sqrt{1-e^2}}{na}\left\{A_r\sin \varphi+A_t\left[\cos \varphi+\rp{1}{e}\left(1-\rp{r}{a}\right)\right]\right\},\\
\dert{i}{t} & = & \rp{1}{na\sqrt{1-e^2}}\ A_n\left(\rp{r}{a}\right)\cos (\omega+\varphi),\lb{in}\\
\dert{\Omega}{t} & = & \rp{1}{na\sin i\sqrt{1-e^2}}\ A_n\left(\rp{r}{a}\right)\sin (\omega+\varphi),\lb{nod}\\
\dert{\omega}{t} & = & -\cos i\dert{\Omega}{t}+\rp{\sqrt{1-e^2}}{nae}\left[-A_r\cos \varphi+A_t\left(1+\rp{r}{p}\right)\sin \varphi\right],\lb{perigeo}\\
\dert{\mathcal{M}}{t} & = & n -\rp{2}{na}\
A_r\left(\rp{r}{a}\right)-\sqrt{1-e^2}\left(\dert{\omega}{t}+\cos
i\dert{\Omega}{t}\right),\lb{manom} \ear
\end{widetext}
in which $n=2\pi/P$ is the  mean motion \footnote{For an unperturbed Keplerian ellipse it amounts to
$n=\sqrt{Gm/a^3}$.}, $P$ is the test particle's orbital period, $\varphi$ is the true anomaly counted from the
pericentre, $p=a(1-e^2)$ is the semilatus rectum of the Keplerian ellipse, $A_r,\ A_t,\ A_n$ are the  radial,
transverse (in-plane components) and  normal (out-of-plane component) projections of the perturbing acceleration
$\boldsymbol{A}$, respectively, on the frame
$\{\boldsymbol{\hat{r}},\boldsymbol{\hat{t}},\boldsymbol{\hat{n}}\}$ comoving with the particle. In our case,
the perturbing acceleration (\ref{ADM}) is entirely radial and, consequently, the Gauss equations reduce to

\bar
\dert{a}{t} & = & \rp{2e}{n\sqrt{1-e^2}}\ A_r\sin \varphi,\lb{smax2}\\
\dert{e}{t} & = & \rp{\sqrt{1-e^2}}{na}\ A_r\sin \varphi,\\
\dert{i}{t} & = & 0,\lb{in2}\\
\dert{\Omega}{t} & = & 0,\lb{nod2}\\
\dert{\omega}{t} & = & -\rp{\sqrt{1-e^2}}{nae}\ A_r\cos \varphi,\lb{perigeo2}\\
\dert{\mathcal{M}}{t} & = & n -\rp{2}{na}\
A_r\left(\rp{r}{a}\right)-\sqrt{1-e^2}\
\dert{\omega}{t},\lb{manom2} \ear

As a result, the inclination and the node are not perturbed by
such a perturbing acceleration.

By evaluating \rfr{ADM} on the unperturbed Keplerian ellipse \eqi r=\rp{a(1-e^2)}{1+e\cos \varphi},\eqf
inserting it into \rfrs{smax2}{manom2} and averaging them with \eqi
\rp{dt}{P}=\rp{1}{2\pi}\rp{(1-e^2)^{3/2}}{(1+e\cos \varphi)^2}d\varphi \eqf one gets that only the argument of
pericentre and the mean anomaly are affected by extra secular precessions \bar \left\langle\dert\omega t
\right\rangle &=&-\rp{3\kappa}{n}\sqrt{1-e^2},\label{dodt}\\
 \left\langle\dert{\mathcal{M}} t
\right\rangle &=&\rp{3\kappa }{n}\left(\rp{7}{3}+e^2\right).\ear As a
consequence, also the mean longitude
$\lambda=\Omega+\omega+\mathcal{M}$, which is used for orbits with
small inclinations and eccentricities, as those of the Solar System
planets, is affected by an extra secular rate
\eqi\left\langle\dert\lambda t
\right\rangle=\rp{3\kappa}{n}\left(\rp{7}{3}+e^2-\sqrt{1-e^2}\right).\label{dldt}\eqf

\section{Comparison with the latest data} \label{sec:conf}

The  expressions obtained are useful for comparison with the latest data on planetary orbits from the EPM2004
ephemerides \cite{pitSSR} by exploiting the determined extra-advances $\Delta\dot\varpi$ of the longitudes of
perihelia $\varpi$ of the inner planets (Table 3 of \cite{pitAL}, here partly reproduced in Table \ref{pitab}).
{\small\begin{table}\caption{ Determined extra-precessions $\Delta\dot\varpi_{\rm obs}$ of the longitudes of
perihelia of the inner planets, in arcseconds per century, by using EPM2004 with the default values
$\beta=\gamma=1$, $J_2^{\odot}=2\times 10^{-7}$. The gravitomagnetic force was not included in the adopted
dynamical force models. Data taken from Table 3 of \cite{pitAL}. It is important to note that the quoted
uncertainties are not the mere formal, statistical errors but are realistic in the sense that they were obtained
from comparison of many different solutions with different sets of parameters and observations (Pitjeva, private
communication 2005a). The correlations among such determined planetary perihelia rates are very low with a
maximum of about $20\%$ between Mercury and the Earth (Pitjeva, private communication 2005b). }\label{pitab}

\begin{tabular}{llll}
\noalign{\hrule height 1.5pt}

 Mercury & Venus  & Earth & Mars\\
$-0.0036\pm 0.0050$ & $0.53\pm 0.30$ & $-0.0002\pm 0.0004$ & $0.0001\pm 0.0005$\\
\hline

\noalign{\hrule height 1.5pt}
\end{tabular}

\end{table}}

The extra-advances $\Delta\dot\varpi$ determined in \cite{pitAL} are affected, in general, by all the Newtonian
and non-Newtonian features of motion which have not been accounted for in the dynamical force models of EPM2004.
Among them, there are certainly the totally unmodelled general relativistic gravitomagnetic field, which induces
the Lense-Thirring planetary precessions, and the solar quadrupole mass moment $J_2^{\odot}$ which, instead, was
included in EPM2004, but it is currently affected by a $\sim 10\%$ uncertainty. For a discussion of such issues
in a different context see \cite{iorlt}. Each planetary perihelion is, thus, affected by such residual effects,
so that it is not possible to entirely attribute the determined extra-advances to the action of $\Delta U$ of
\rfr{poti}, especially for Mercury for which the mismodelled/unmodelled Newtonian and Einsteinian perturbations
are stronger. A better way to use $\Delta\dot\varpi$ is to suitably combine them in order to make the estimate
of $\kappa$ independent, by construction, of the Lense-Thirring and $J_2^{\odot}$ effects. According to the
approach followed in \cite{iorlt, ioraa}, it is possible to construct the following combination \eqi
\kappa=\rp{\Delta\dot\varpi^{\rm Mer}+b_1\Delta\dot\varpi^{\rm Ear}+b_2\Delta\dot\varpi^{\rm Mar}}{-1.97\times
10^{9}\ {\rm s}} ,\lb{combi1}\eqf with $b_1=-80.7$ and $b_2=217.6$. The dimensionless coefficients $b_1$ and
$b_2$, which are built up with $a,e$ and $i$ of the planets adopted, cancel out, by construction, the impact of
the gravitomagnetic field and of the solar quadrupolar mass moment on the combination \rfr{combi1}. This can
straightforwardly be checked by combining the perihelion precessions due to $J_2^{\odot}$ and to the
Lense-Thirring force with the coefficients of \rfr{combi1}: the result is zero. \Rfr{combi1} and the values of
Table \ref{pitab}, converted in s$^{-1}$, allow to obtain \eqi \kappa=-2.6\times 10^{-26}\ {\rm s}^{-2.}
\label{eq:estkappa1} \eqf The error on $\kappa$ obtainable from \rfr{combi1} can conservatively be evaluated as
\footnote{Although rather small, a certain amount of correlation among the determined extra-advances of
perihelia is present, with a maximum of $20\%$ between the Earth and Mars (E.V. Pitjeva, private communication,
2006). If a root-sum-square calculation is performed, an upper bound $\kappa= 8\times 10^{-26}$ s$^{-2}$ is
obtained.}
\begin{align}
\delta \kappa& \leq \rp{\delta(\Delta\dot\varpi^{\rm
Mer})+|b_1|\delta(\Delta\dot\varpi^{\rm
Ear})+|b_2|\delta(\Delta\dot\varpi^{\rm Mar})}{ 1.9\times 10^{9}\
{\rm s} } \notag \\ &=1\times 10^{-25}\ {\rm s}^{-2}.
\end{align}

\section{Discussion} \label{sec:disc}

Because of the relations  $k=3\kappa$, and $k=R/4$, from the estimate (\ref{eq:estkappa1}) on $\kappa$,  we get corresponding bounds on the allowed
values of the scalar curvature $R$:

\begin{equation}
|R| \leq 3.12 \times 10^{-25}  {\rm s}^{-2}, \label{eq:stimaR1}
\end{equation}
or, restoring units in such a way that $R$ is measured in $m^{-2}$, we have

\begin{equation}
|R| \leq 3.47 \times 10^{-42} {\rm m}^{-2}. \label{eq:stimaR2}
\end{equation}

Since the $k$ parameter of the Schwarzschild-de Sitter metric can be interpreted as a cosmological constant
$\Lambda$ (see \cite{multamaki06}), we get for $\Lambda$ the following limits

\begin{equation}
|k|=|\Lambda| \leq 8.68 \times 10^{-43} {\rm m}^{-2}. \label{eq:stimaR3}
\end{equation}
This result is a much smaller than the one obtained in \cite{kagramanova06}, and comparable to the those
obtained in \cite{sereno06a} and \cite{ioriodm}, but it is still several orders of magnitude greater than the
current value of the cosmological constant $\Lambda_0 \simeq 10 ^{-52}  {\rm m}^{-2}$ \cite{peebles03}.

As we have seen in Sec. \ref{sec:frgravity}, the $k$ parameter is simply related to the solutions of the
structural equation
\begin{equation}
f^{\prime }(R) R-2f(R)= 0, \label{eq:struct1}
\end{equation}
namely it is $k=c_i/4$, where $R=c_i$ are the solutions of eq. (\ref{eq:struct1}). As a consequence, the bounds
on the $\kappa$ parameter, in principle,  enable us to constrain the functions $f(R)$. We may proceed as
follows.

In general, the functions $f$, beyond the scalar curvature $R$, depend on a set of $N$ constant parameters
$\alpha_j$, $j=1..N$, so that we may write $f=f(R,\alpha_1,..,\alpha_N)$, and, on solving eq.
(\ref{eq:struct1}), we obtain
\begin{equation}
R=\mathcal{F}(\alpha_1,..,\alpha_N). \label{eq:falpha1}
\end{equation}

Consequently, what we ultimately obtain is a limit on the allowed values of the combination $\mathcal{F}$ of
these parameters.

In \cite{c6} the Lagrangian
\begin{equation}
f(R)=R-\frac{\mu^4}{R} \label{eq:frcarrol1}
\end{equation}
was introduced, and it was proved that it mimics cosmic acceleration without need for dark energy. However, we
point out that the Lagrangian (\ref{eq:frcarrol1}) was found to be instable \cite{metfr1,metfr2,dolgov03}. For
discussion on this issue, we refer to the recent paper by Faraoni \cite{faraoni06b}. Nonetheless, we consider
the Lagrangian (\ref{eq:frcarrol1}) as an example in order to illustrate our approach, while a more general
discussion will be carried out elsewhere \cite{allemandi06}.

The Lagrangian (\ref{eq:frcarrol1}) depends on the parameter $\mu$ only, and eq. (\ref{eq:falpha1}) becomes
\begin{equation}
|R|=\sqrt{3}\mu^2,  \label{eq:frcarrol2}
\end{equation}
and, thanks to (\ref{eq:stimaR3}), we may set a limit on the parameter $\mu$:
\begin{equation}
\mu \leq 1.41 \times 10 ^ {-21}  {\rm m}^{-1}, \label{eq:carrol3}
\end{equation}
or, since $\mu$ is a parameter whose physical units are those of a mass,
\begin{equation}
\mu \leq 2.80 \times 10 ^ {-28}  {\rm eV}. \label{eq:carrol4}
\end{equation}
This value is remarkably greater than estimate $\mu \simeq 10 ^ {-33} \  {\rm eV}$ \cite{c6}, needed for $f(R)$
gravity to explain the acceleration of the Universe without requiring dark matter.

The same result hold for the slightly different Lagrangian $f(R)=R-\frac{\mu^4}{R}+\alpha R^2$.\\

The same approach that we have briefly outlined here, can be applied to other functions $f(R)$, and it could be useful, in
particular, for those for which the structural equation can be solved analytically.

\section{Conclusions} \label{sec:conc}

In this paper, we have studied the $f(R)$ Higher Order Theories of Gravity and, in particular, the effects of
the modifications of the Newtonian potential   on the Keplerian orbital elements of the Solar System planets.
Starting from an exact solution of the field equations in vacuum in the Palatini formalism, which corresponds to
the Schwarzschild-de Sitter metric, we have considered the perturbations of the orbital elements, by means of
the Gauss equations. We have showed that the  longitudes of perihelia  and the mean longitudes  are affected by
secular precessions and, by comparison with the latest results in planetary orbit determination from the EPM2004
ephemerides, we have obtained the resulting best estimate of the parameter $k$ which measures the non linearity
of the gravitational theory. Since the $k$ parameter can be interpreted as a cosmological constant $\Lambda$,
our approach enables us to set the limit $|\Lambda| \leq 8.68 \times 10^{-43}  {\rm m}^{-2}$, which is  much
smaller  than the one obtained in \cite{kagramanova06} and comparable to the those obtained in
\cite{sereno06a,ioriodm}, but still too big if confronted with the current value of the cosmological constant
$\Lambda_0 \simeq 10 ^{-52} {\rm m}^{-2}$. Furthermore, we have showed that the best estimate of the parameter
$k$, can be used to constrain the functions $f(R)$. In particular, our approach is suitable for those $f(R)$ for
which the structural equation (\ref{eq:struct1}) can be solved analytically, thus allowing an explicit
evaluation of the parameters appearing in them from cosmological experiments results. As an example, we have
considered $f(R)=R-\mu^4/R$, and we have obtained the limit for the parameter $\mu:\mu \leq 2.80 \times 10 ^
{-28}  {\rm eV}$, much greater than  $\mu \simeq 10 ^ {-33} \  {\rm eV}$, required in order to agree with
cosmological observations  \cite{c6}.

What our results, in agreement with \cite{sereno06a,sereno06b,kagramanova06,multamaki06}, suggest is that, in
general, Solar System experiments are not able to constrain $k$ or the other related parameters up to orders of
magnitude comparable to the cosmologically relevant values. On the other hand, this fact can be also interpreted
by saying that $f(R)$ theories are viable on the Solar System scale, since their predictions are
indistinguishable from GR ones, while they are appreciably different only on much larger scales, such as the
cosmological one.

All we have done, refers to a solution of the field equation in vacuum, and corresponds to a space-time metric
of constant scalar curvature $R$. Actually, a complete study of the comparison between GR and $f(R)$ theories
needs more general situations, such as, for instance, those requiring solutions of the field equations within
the matter distribution: this will be done elsewhere.

\section*{ACKNOWLEDGMENTS}

The authors  are very grateful to Prof. A. Tartaglia and Dr. G. Allemandi for helpful comments and suggestions.
M.L.R  acknowledges financial support from the Italian Ministry of University and Research (MIUR) under the
national program 'Cofin 2005' - \textit{La pulsar doppia e oltre: verso una nuova era della ricerca sulle
pulsar}.


\end{document}